\documentclass[fleqn,10pt]{wlscirep}
\usepackage[utf8]{inputenc}
\usepackage[T1]{fontenc}
\usepackage{makecell}
\usepackage{nicematrix}
\usepackage{enumitem}
\usepackage{amsthm}
\usepackage{amssymb }
\usepackage{bm}
\usepackage{cite}
\usepackage{float}
\usepackage{mathrsfs}
\usepackage{amsfonts}
\usepackage{bm}
\usepackage{float}
\usepackage{mathrsfs}
\usepackage{graphicx}
\usepackage{algorithm, algpseudocode}
\usepackage{subfigure}
\usepackage{tabularx}
\usepackage{flushend}
\usepackage{epsfig}
\usepackage{cite}
\usepackage[super]{nth}
\usepackage{flushend}
\usepackage{multirow}
\usepackage{color}
\usepackage{enumerate}
\usepackage{easylist}
\usepackage{kantlipsum}

\def\MM{\text{CT-HGR}}
\def\p{p}
\def\x{{\bm{x}}}
\def\xp{\x^{\p}}
\def\E{\bm{E}}

\def\z{z}
\def\N{N}
\def\Q{Q}
\def\K{K}
\def\V{V}
\def\d{d}
\def\l{l}
\def\L{L}
\def\i{i}
\def\y{y}

\title{Transformer-based Hand Gesture Recognition via High-Density EMG Signals: From Instantaneous Recognition to Fusion of Motor Unit Spike Trains}

\author[1]{Mansooreh Montazerin}
\author[2]{Elahe Rahimian}
\author[2]{Farnoosh Naderkhani}
\author[3,4]{S. Farokh Atashzar}
\author[5]{Svetlana Yanushkevich}
\author[1,2,*]{Arash Mohammadi}
\affil[1]{Department of Electrical and Computer Engineering, Concordia University, Montreal, QC, Canada}
\affil[2]{Concordia Institute for Information Systems Engineering, Concordia University, Montreal, QC, Canada}
\affil[3]{New York University (NYU), Departments of Electrical \& Computer Engineering and Mechanical \& Aerospace Engineering, New York, 10003, NY, USA}
\affil[4]{New York University (NYU), NYU WIRELESS, and NYU Center for Urban Science and Progress (CUSP), New York, 10003, NY, USA}
\affil[5]{Department of Electrical and Software Engineering, Biometric Technologies Laboratory, Schulich School of Engineering, University of Calgary, Calgary, AB, Canada}
\affil[*]{arash.mohammadi@concordia.ca}

\begin{abstract}
Designing efficient and labor-saving prosthetic hands requires powerful hand gesture recognition algorithms that can achieve high accuracy with limited complexity and latency. In this context, the paper proposes a compact deep learning framework referred to as the $\MM$, which employs a vision transformer network to conduct hand gesture recognition using high-density sEMG (HD-sEMG) signals. Taking advantage of the attention mechanism, which is incorporated into the transformer architectures, our proposed $\MM$ framework overcomes major constraints associated with most of the existing deep learning models such as model complexity;  requiring feature engineering; inability to consider both temporal and spatial information of HD-sEMG signals, and; requiring a large number of training samples. The attention mechanism in the proposed model identifies similarities among different data segments with a greater capacity for parallel computations and addresses the memory limitation problems while dealing with inputs of large sequence lengths. $\MM$ can be trained from scratch without any need for transfer learning and can simultaneously extract both temporal and spatial features of HD-sEMG data. Additionally, the $\MM$ framework can perform instantaneous recognition using sEMG image spatially composed from HD-sEMG signals. A variant of the $\MM$ is also designed to incorporate microscopic neural drive information in the form of Motor Unit Spike Trains (MUSTs) extracted from HD-sEMG signals using Blind Source Separation (BSS). This variant is combined with its baseline version via a hybrid architecture to evaluate potentials of fusing macroscopic and microscopic neural drive information. The utilized HD-sEMG dataset involves $128$ electrodes that collect the signals related to $65$ isometric hand gestures of $20$ subjects. The proposed $\MM$ framework is applied to $31.25$, $62.5$, $125$, $250$ ms window sizes of the above-mentioned dataset utilizing $32$, $64$, $128$ electrode channels. The average accuracy over all the participants using $32$ electrodes and a window size of $31.25$ ms is $86.23$\%, which gradually increases till reaching $91.98$\% for $128$ electrodes and a window size of $250$ ms. The $\MM$ achieves accuracy of $89.13$\% for instantaneous recognition based on a single frame of HD-sEMG image. The proposed model is statistically compared with a 3D Convolutional Neural Network (CNN) and a Support Vector Machine (SVM) model. The results corroborate effectiveness of the proposed framework compared to its counterparts.
\end{abstract}

\begin{document}
\flushbottom
\maketitle
\thispagestyle{empty}

\section{Introduction}
Hand gesture recognition using surface Electromyogram (sEMG) signals has been a topic of growing interest for development of assistive systems to help individuals with amputated limbs. Generally speaking, myoelectric prosthetic devices work by classifying existing patterns of the collected sEMG signals and synthesizing the intended gestures~\cite{li2021gesture}. While conventional myoelectric control systems, e.g., on/off control or direct-proportional, have potential advantages, challenges  such as limited Degree of Freedom (DoF) due to crosstalk have resulted in the emergence of data-driven solutions. More specifically, to improve efficiency, intuitiveness, and the control performance of hand prosthetic systems, several Artificial Intelligence (AI) algorithms ranging from conventional Machine Learning (ML) models to highly complicated Deep Neural Network (DNN) architectures have been designed for sEMG-based hand gesture recognition in myoelectric prosthetic devices~\cite{rahimian2021,rahimianhand,farina2021, tam2021}. The ML-based models encompass traditional approaches such as  Support Vector Machines (SVMs), Linear Discriminant Analysis (LDA), and $k$-Nearest Neighbors (kNNs)~\cite{chen2019hand, svm2021,lda2019,ml2022}, and DNN-based models consist of frameworks such as Convolutional Neural Networks (CNNs), Recurrent Neural Networks (RNNs), and Transformer-based architectures~\cite{ann2021,dilated2019,chen2020,azhiri2021real,simao2019emg,temgnet2021}. 

sEMG signals represent the electrical activities of the muscles and are recorded by a set of non-invasive electrodes that are placed on the muscle tissue~\cite{toledo2018,jiang2012}. Broadly speaking, there are two types of sEMG acquisition systems, called sparse and high-density~\cite{HD2021,sparse2019}. Both of these groups are obtained by placing electrodes on the surface of the muscle and recording the electrical activity of the muscle's motor unit action potentials in response to the neural signals. Unlike sparse sEMG acquisition that involves a limited number of electrodes to record muscle activities, High-density sEMG (HD-sEMG) signals are obtained through a two-dimensional (2D) grid of electrodes, which cover an area of the muscle tissue and a large number of associated motor units~\cite{rojashd,baihd}. It is, therefore, more difficult to design an ML/DL-based algorithm for hand gesture recognition from HD-sEMG signals as they require more computational power for the signal processing and training stages. However, HD-sEMG signals are considered more potent than their sparse counterparts because of their ability to include both temporal and spatial information of muscle activities, which provides a high-resolution 3-dimensional signal (two dimensions in space and one in time)~\cite{rojas2020}. The HD-sEMG signal acquisition can evaluate functionality of the underlying neuromuscular system more precisely in terms of spatial resolution. Accordingly, developing an efficient DNN-based framework that can effectively learn from a comprehensive HD-sEMG dataset is of great importance in neuro-rehabilitation research and clinical trials~\cite{campanini2020}, which is the focus of this manuscript.

Conventional ML models, such as SVMs and LDAs, utilized for sEMG-based hand gesture recognition, typically work well when dealing with small datasets. These methods, however, depend on manual extraction of handcrafted (engineered) features, which limits their generalizability as human knowledge is needed to find the best set of features~\cite{yang2021}. Increasing the number of utilized electrodes and the number of gestures entails extracting more features, therefore, the feature extraction process becomes significantly complex and time-consuming. This is because more trials and efforts are required to boost the discriminative power of the model. Dependence on engineered features is partially/fully relaxed by utilization of DNN-based models. Among the most frequently used DNN architectures for the task of hand gesture recognition is the CNN-based frameworks. For example, Reference~\cite{chen2020} converts sEMG signals to 3D images and uses transfer learning to feed them to a popular CNN trained on a database of natural images. CNNs, however, are designed to concentrate on learning spatial features of the input signals and fail to extract temporal features of the sEMG data. To overcome this issue, researchers turned their attention to hybrid CNN-RNN frameworks that were designed to take both spatial and temporal information of the time-series sEMG datasets into account~\cite{hybrid2018,hybrid2022}. For instance, Hu \textit{et~al.}~\cite{hybrid2018} have applied attention mechanism on top of a hybrid CNN-LSTM model to perform hand gesture recognition based on sEMG signals with relatively large window sizes (i.e. $150$ ms and $200$ ms). They achieved classification accuracy of up to $87\%$ using the largest window size. In~\cite{hybrid2022}, a dimensionality reduction method is proposed and assumed to enhance the classification accuracy when used with a hybrid CNN-LSTM architecture. In this framework~\cite{hybrid2022}, the classification accuracy is  $88.9\%$ on the same dataset as that of~\cite{hybrid2018} for the $250$ ms window size. Nonetheless, as well as not allowing entire input parallelization, hybrid CNN-RNN frameworks are usually computationally demanding and reveal important limitations with respect to the memory usage and large training times. In this paper, we aim to eliminate the complexity of simultaneously exploiting CNNs and RNNs by introducing a Vision Transformer-based (ViT)~\cite{ViT} architecture to be applied on HD-sEMG signals and to efficiently deal with the above-mentioned constraints.

In this study, a comprehensive evaluation of the proposed ViT-based framework for hand gesture classification on HD-sEMG dataset is carried out for the first time to the best of our knowledge. The ViT architecture takes advantage of the attention~\cite{attention} mechanism, which works by finding dependencies and similarities among different data portions. The attention mechanism in the ViT is integrated in a typical transformer model, making it a robust framework for hand gesture recognition without being combined with other DL algorithms. One of the differences between the ViT and a typical transformer is that the ViT is generally designed to be applied on 3D images rather than 2D time-series signals. Considering the fact that HD-sEMG signals comprise of two dimensions in space and one in time (3 dimensions in total), they can be an appropriate input to a ViT. As mentioned in~\cite{butter}, instantaneous training with HD-sEMG signals refers to training the network with a 2D image depicting motor unit action potential activities under a grid of electrodes at a single time point. In this paper, we also show that there are reproducible patterns among instantaneous samples of a specific gesture which could also be a physiological representation of muscle activities in each time point. We demonstrate that the proposed framework can perform instantaneous hand gesture classification using sEMG image spatially composed from HD-sEM. In other words, it can achieve acceptable accuracy when receiving, as an input, a single frame of the HD-sEMG image. The main contributions of the paper are briefly outlined below:
\begin{itemize}[noitemsep]
\item To the best of our knowledge, the proposed $\MM$ is the first ViT-based architecture that is leveraged to classify hand gestures from HD-sEMG signals. It can efficiently classify a large number of hand gestures relying only on the attention mechanism. Furthermore, the $\MM$ can be trained from scratch without the need for transfer learning or data augmentation.
\item Achieving high accuracy over small windows sizes, e.g., $32$ ms, which has been rarely worked on in the previous literature. This can be a remarkable milestone for hand prosthetic systems to operate with low latency.
\item Achieving near baseline accuracy using instantaneous HD-sEMG data samples, which is significant  as it paves the way for real-time learning from HD-sEMG signals.
\item Introducing, for the first time to the best of our knowledge, the idea of integrating macroscopic and microscopic neural drive information through a hybrid DNN framework. The proposed variant of the $\MM$ framework,  is a hybrid model that simultaneously extracts a set of temporal and spatial features through its two independent ViT-based parallel architectures (the so called Macro and Micro paths). The Macro Path is the baseline $\MM$ model, while the Micro path is fed with the p-to-p values of the extracted Motor Unit Action Potentials (MUAPs) of each source.
\end{itemize}
The rest of the paper is structured as follows: The utilized HD-sEMG dataset is introduced in Sub-Section~\ref{sub:data}. An explanation of the pre-processing procedures on the raw dataset is given in Sub-section~\ref{sub:prepro} and our proposed framework is presented in Sub-section~\ref{sub:frame}. Our experiments and evaluations of implementing the proposed framework are discussed in Section~\ref{sec:results} and finally, Section~\ref{sec:discuss} concludes the paper.

\section{Materials and Methods} \label{sec:materials}

\subsection{The HD-sEMG Dataset} \label{sub:data}

\begin{figure}[t!]
\centering
\includegraphics[width=0.5\linewidth]{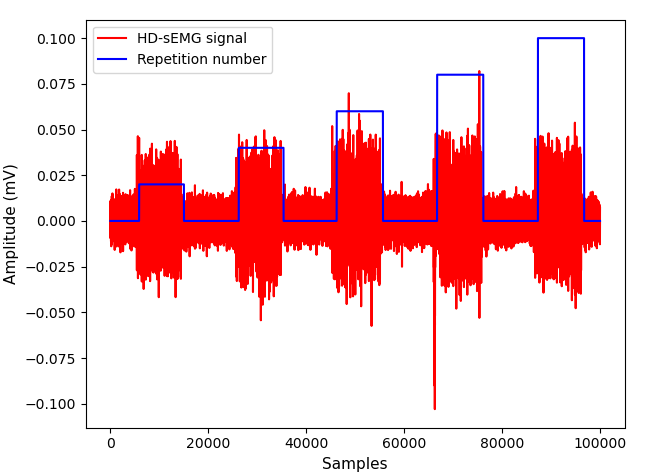}
\caption{\small Illustration of the raw HD-sEMG dataset. The red plot is the sEMG signal and the blue plot shows the exact location of movements and their repetition number.}
\label{raw}
\end{figure}

The dataset~\cite{data} used in this study is a recently released HD-sEMG dataset that contains two $64$-electrode square grids ($8\times 8$) with an inter-electrode distance of $10$ mm, which were placed on extensor and flexor muscles of $20$ participants. One of the subjects is not included in the study from the beginning due to its incomplete information. The participants performed $65$ hand gestures that are combinations of $16$ basic single degree of freedom movements. One of the gestures is carried out twice, therefore, there are $66$ movements in total. The subjects performed each gesture $5$ times with $5$ seconds rest in between. Fig.~\ref{raw} illustrates how the raw dataset is organized. The red plot shows the HD-sEMG signal acquired for all the $5$ repetitions of one specific hand movement. The blue line shows the repetition number of the gesture and the rest intervals. The signals were recorded through a Quattrocento (OT Bioelettronica, Torino, Italy) bioelectrical amplifier system with $2,048$ Hz sampling frequency. Signals of the successive channels were subtracted from each other to lower the amount of common-mode noise. The rational behind selection of this publicly available dataset is that it comprises of a large number of gestures and electrodes, which allows development of a generalizable framework by investigating different settings of the input data. Additionally, this dataset provides straightforward instructions on how to deploy the dataset for different evaluation purposes. However, since the paper~\cite{data} on this dataset did not refer to the train and test sets as a basis for comparison, we performed a $5$-fold cross-validation as there are $5$ sessions in the dataset. In this way, one (out of $5$) repetition is considered as the test set and the remaining are assigned to the train set. Each time, the test set is changed until all the repetitions have been tested. Finally, the accuracy of each fold together with the average accuracy across all the folds are reported.

\subsection{Data Pre-processing}\label{sub:prepro}
The raw HD-sEMG dataset is pre-processed in three distinct steps before being fed to the proposed $\MM$ framework. More specifically, the pre-processing steps in this study consist of low-pass filtering, normalization, and windowing. During hand movement recording, sEMG signals were filtered with a $10$-Hz high-pass.
However, following several experiments and based on the previous literature~\cite{mansoorehViT,butter,atzorifilter,zhangfilter}, we deduced that the wide variations in the sEMG signals prevent the model from learning key features properly. Thus, a low-pass first-order butterworth filter at $1$ Hz is applied separately to each of the $128$ channels of the data to derive the positive envelope of the main signal. In the next pre-processing phase, the filtered signals are normalized by the $\mu$-law normalization algorithm, which reduces significant changes in the dynamic range of the signals acquired from different electrodes. The $\mu$-law normalization is performed based on the following formulation
\begin{equation}\label{mu_law}
F(x_t) = \text{sign}(x_t)\frac{\ln{\big(1+ \mu |x_t|\big)}}{\ln{\big(1+ \mu \big)}},
\end{equation}
where $x_t$ is the time-series sEMG signal for each electrode channel, and $\mu$ is the extent to which the signals are scaled down and is determined empirically. According to~\cite{IcasspElahe,rahimian2021}, $\mu$-law normalization helps the network to learn gestures more effectively. The final pre-processing stage is to segment the sEMG signals. After removing the rest intervals from the dataset, the signals are segmented with a specific window size creating the main 3D input of the $\MM$ with shape $W\times N_{ch}\times N_{cv}$, where $W$ is the window size and $N_{ch}$ and $N_{cv}$ are the number of horizontal and vertical channels respectively. This completes our discussion on the pre-processing stage. In what follows, the proposed $\MM$ framework is presented, which takes the pre-processed data samples as its input and returns the predicted gesture class.

\subsection{The Proposed \texorpdfstring{$\MM$} {Framework}}\label{sub:frame}
In this section, description of our proposed $\MM$ framework, its main building blocks, and its adoption for the task of hand gesture recognition are presented. The $\MM$ is developed based on the ViT network in which the attention mechanism is utilized to understand the temporal and spatial connections among multiple data segments of the input. As stated previously, several studies have employed the attention mechanism together with hybrid CNN-RNN models to force the network to learn both spatial and temporal information of the signals~\cite{hybrid2018,rahimianhand}. However, in this paper, we demonstrate that attention mechanism can work independently of any other network and achieve high accuracy when trained from scratch with no data augmentation. We also show that the proposed framework can be trained even on pretty small window sizes and more importantly on  instantaneous data samples.

\begin{figure}[t!]
\centering
\includegraphics[width=0.7\linewidth]{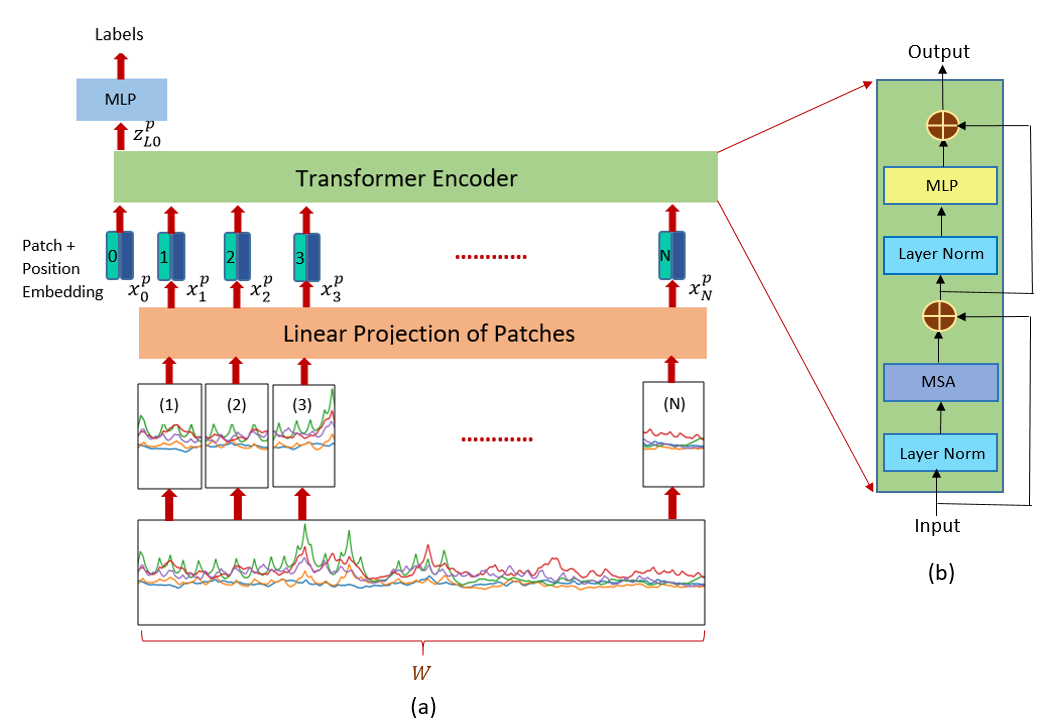}
\caption{\footnotesize Overview of the $\MM$ network. (a) The windowed HD-sEMG signal is fed to the $\MM$ and split into smaller patches. The patches go through a linear projection layer which converts them from 3D to 2D data samples. A class token is added to the patches and the $N+1$ patches are input to a transformer encoder. Ultimately, the first output of the transformer corresponding to the class token is chosen for the multi-class classification part. (b) This is the transformer encoder which is the fundamental part of the ViT, responsible for processing the input patches.}
\label{ViT}
\end{figure}

An overall illustration of the $\MM$ is indicated in Fig.~\ref{ViT}. After completion of the pre-processing steps discussed in the previous section, we have 3D signals of shape $W\times N_{ch}\times N_{cv}$, where $W$ is the window size and $N_{ch}$ and $N_{cv}$ are the number of horizontal and vertical channels respectively. The utilized window sizes in our experiments are mainly of $64$, $128$, $256$, and $512$ data points ($31.25$, $62.5$, $125$, and $250$ ms respectively). We choose the window sizes to be the powers of $2$ because it provides us with more flexibility in patching the data segments after being fed to the ViT~\cite{winjustify}. Furthermore, we have assessed the effect of changing the number of electrode channels by using $32$, $64$ and $128$ out of the whole $128$ channels. Therefore, we set $N_{ch}$ to $4$, $8$, and $16$ each time while $N_{cv}$ remains constant at $8$. In what follows, the major blocks of the proposed $\MM$ network, namely ``Patch Embedding'', ``Position Embedding'', ``Transformer Encoder'', and the ``Multilayer Perceptron'' blocks.

\subsubsection{Patch Embedding}\label{sub:patch}
In this block, the 3D signals are divided into $N$ small patches either horizontally, vertically or both. Therefore, we have $N$ patches of size $H\times V\times N_{cv}$ that are then linearly flattened to 2D signals of size $N\times HVN_{cv}$ where, $N$ is equal to $WN_{ch}/HV$ and is the effective sequence length of the transformer's input. Consequently, there are $N$ patch vectors $\xp_{i}$, for ($1 \leq i \leq N$). Using a trainable linear projection layer, the $\xp_{i}$ vectors are embedded with the model's dimension $d$. The linear projection is shown with matrix $\bm{E}$, which is multiplied to each of the $\xp_{i}$ and yields $N$ vectors of dimension $d$. Moreover, a class token named $\xp_{0}$ similar to what was previously used in the Bert framework~\cite{bert} is prepended to the aforementioned vectors to gather all the useful information learned during the training stage and is used in the final step when different hand gestures are classified. The final sequence length of the transformer after adding the class token is $N+1$.

\subsubsection{Position Embedding}\label{sub:pos}
Unlike RNNs that process their inputs sequentially, transformers apply the attention mechanism to all of the data segments in parallel, which deprives them of the capacity to intrinsically learn about the relative position of each patch of a single input. Because sEMG signals are time-series sequences of data points in which the location of each point matters for hand gesture classification tasks, we need to train the network to assign a specific position to each sample. Generally speaking, positional embedding is an additional piece of information that is injected into the network, helping it to identify how data points are ordered. There are different types of positional embeddings offered such as relative, 1D, 2D, and sinusoidal positional embeddings that may be learnable or non-learnable. In this context, we use a learnable 1D positional embedding vector that is added to each of the embedded $\xp_{i}$ vectors to maintain and learn the position of each patch during the training phase.
The final output $\z_0$ of the ``Patch + Position Embedding'' blocks is given by
\begin{eqnarray}
\centering
\z_0 = [\xp_0; \xp_1\E; \xp_2\E;\dots; \xp_\N\E] + \E^{pos},
\end{eqnarray}
where $\E^{pos}$ is an $(\N+1)\times d$ matrix, holding the relative position of each patch in a $d$-dimensional vector.

\subsubsection{Transformer Encoder}\label{sub:encoder}
A typical transformer model consists of two major parts called encoder and decoder. In this paper, we aim to utilize only the former part. The transformer encoder is where the attention mechanism tries to find the similarities among the $N+1$ patches that arrive at its input. As can be seen in Fig.~\ref{ViT}(b), there are $L$ identical layers of transformer encoder in the $\MM$ network and each has three separate blocks, named as ``Layer Norm'', ``Multi-head Self Attention (MSA)'' and ``MLP''. The $\z_0$ sequence of patches that is explained above is first fed to a normalization layer to improve the generalization performance of the model and accelerate the training process~\cite{norm}. The ``Layer Norm block'' is then followed by the MSA module, which incorporates $h$ parallel blocks (heads) of the scaled dot-product attention (also known as self attention). In the context of self attention, three different vectors $Keys (K)$, $Queries (Q)$ and $Values (V)$ of dimension $d$ are employed for each input patch. For computing the self attention metric, the dot product of $Queries$ and all the $Keys$ are calculated and scaled by 1/$\sqrt{d}$ in order to prevent the dot products from generating very large numbers. This matrix is then, converted into a probability matrix through a $softmax$ function and is multiplied to the $Values$ to produce the attention metric as follows
\begin{eqnarray}
Attention = \text{$softmax$}(\frac{\Q\K^T}{\sqrt{\d}})\V. \label{eq.3}
\end{eqnarray}
In the MSA block, instead of dealing with $d$-dimensional $Queries$, $Keys$ and $Values$, we split them into $h$ parallel heads and measure the self attention metric on these heads independently. Finally, after finding the corresponding results for each head, we concatenate them to obtain the $d$-dimensional vectors of patches. As indicated in Fig.~\ref{ViT}(b), residual paths from the encoder's input to the output of the MSA block are employed to avoid the gradient vanishing problem. The formulations for the above explanations are as follows
\begin{eqnarray}
\z^{'}_l &=& MSA(LayerNorm(\z_{\l-1})) + \z_{\l-1},\label{eq:MSA}\\
\z_l &=& MLP(LayerNorm(\z^{'}_{\l})) + \z^{'}_{\l}, \label{eq:MLP}
\end{eqnarray}
where $\z_l$ is the $l^{th}$ transformer layer's output and $l=1,\dots, \L$. The final output of the transformer encoder is given by
\begin{eqnarray}
\z_\L = [\z^{\p}_{\L0}; \z^{\p}_{\L1}; \dots; \z^{\p}_{\L\N}],
\end{eqnarray}
where $\z^{\p}_{\L\i}$ is the final layer's output corresponding to the $\i^{th}$ patch and $\i=1,\dots, \N$. As mentioned before, among all the above vector of patches, the $\z^{\p}_{\L0}$ vector matching the class token is chosen for gesture classification. Authors in~\cite{ViT} claim that the learned features in the sequence of patches will eventually be included in the class token, which has a decisive role in predicting the model's output. Therefore, $\z^{\p}_{\L0}$ is passed to a linear layer which outputs the predicted gesture's label as
\begin{eqnarray}
\y_{\text{predicted}} = \text{$Linear$}(\z^{\p}_{\L0}).
\end{eqnarray}
In the next section, we will describe the various experiments performed in this study and present the obtained results and their explanations in detail.

\section{Results}\label{sec:results}
We perform several experiments to evaluate performance of the proposed framework under different configurations. In the following, each of the conducted experiments and their corresponding results are presented separately. The implemented models are evaluated on all the $66$ gestures of the HD-sEMG dataset performed by $19$ healthy subjects. The implementations were developed in the PyTorch framework and the models are trained using an NVIDIA GeForce GTX 1080 Ti GPU.

\subsection{Overall Performance Evaluation under Different Configurations}\label{sub:diff}
\begin{table}[t!]
\begin{center}
\caption{\footnotesize Comparison of classification accuracy and STD for each fold and their average for different window sizes and number of channels ($\MM$-V1). The accuracy and STD for each fold is averaged over $19$ subjects.}
 \label{model1}
\resizebox{\columnwidth}{!}{
 {\begin{tabular}{  c  c  c c c c c c }
\hline
\hline
\textbf{\# Channels}
& \textbf{Window size}
& \textbf{Fold1(\%)}
& \textbf{Fold2}
& \textbf{Fold3}
& \textbf{Fold4}
& \textbf{Fold5}
& \textbf{Average}
\\
\hline
    \multirow{3}{*}{\textbf{32}}
& \textbf{64}
& 76.85 ($\pm$3.83)
& 89.30 ($\pm$2.61)
& 89.91 ($\pm$2.54)
& 89.62 ($\pm$2.67)
& 85.49 ($\pm$3.07)
& 86.23 ($\pm$2.94)
\\
& \textbf{128}
& 77.21 ($\pm$3.56)
& 89.48 ($\pm$2.60)
& 90.05 ($\pm$2.63)
& 90.00 ($\pm$2.61)
& 85.83 ($\pm$2.96)
& 86.51 ($\pm$2.87)
\\
& \textbf{256}
& 77.63 ($\pm$3.50)
& 90.51 ($\pm$2.52)
& 90.79 ($\pm$2.45)
& 90.99 ($\pm$2.42)
& 86.66 ($\pm$2.97)
& 87.32 ($\pm$2.77)
\\
\hline
\multirow{3}{*}{\textbf{64}}
& \textbf{64}
&79.64 ($\pm$3.38)
&91.92 ($\pm$2.41)
&92.55 ($\pm$2.18)
&92.37 ($\pm$2.32)
&88.16 ($\pm$2.77)
&88.93 ($\pm$2.61)
\\
& \textbf{128}
&80.26 ($\pm$3.44)
&92.32 ($\pm$2.27)
&92.94 ($\pm$2.20)
&92.48 ($\pm$2.22)
&88.46 ($\pm$2.77)
&89.29 ($\pm$2.58)
\\
& \textbf{256}
& 81.43 ($\pm$3.31)
& 92.89 ($\pm$2.15)
& 93.42 ($\pm$2.13)
& 93.05 ($\pm$2.18)
& 89.29 ($\pm$2.69)
& 90.02 ($\pm$2.49)
\\
\hline
\multirow{4}{*}{\textbf{128}}
& \textbf{64}
& 82.14 ($\pm$3.26)
&  93.30 ($\pm$2.14)
&  93.75 ($\pm$2.08)
&  93.39 ($\pm$2.11)
&90.07 ($\pm$2.55)
&90.53 ($\pm$2.43)
\\
& \textbf{128}
& 82.80 ($\pm$3.22)
&  93.47 ($\pm$2.13)
&  93.98 ($\pm$2.03)
&  93.82 ($\pm$2.10)
&  90.30 ($\pm$2.48)
& 90.87 ($\pm$2.39)
\\
& \textbf{256}
& 83.20 ($\pm$3.21)
& 94.19 ($\pm$2.00)
& 94.25 ($\pm$1.97)
& 94.42 ($\pm$1.91)
& 90.70 ($\pm$2.46)
& 91.35 ($\pm$2.31)
\\
& \textbf{512}
& 83.87 ($\pm$3.21)
& 94.62 ($\pm$1.88)
& 95.26 ($\pm$1.80)
& 94.89 ($\pm$1.85)
& 91.26 ($\pm$2.37)
& 91.98 ($\pm$2.22)
\\
\hline
\hline
\end{tabular}}
}
\end{center}
\end{table}
\begin{table}[t!]
 \begin{center}
 \caption{\footnotesize Comparison of classification accuracy and STD for each fold and their average for different window sizes and $128$ electrode channels ($\MM$-V2). The accuracy and STD for each fold is averaged over $19$ subjects.}
 \label{model2}
 \resizebox{\columnwidth}{!}{
 {\begin{tabular}{  c  c  c c c c c c}
\hline
\hline
\textbf{\# Channels}
& \textbf{Window size}
& \textbf{Fold1(\%)}
& \textbf{Fold2}
& \textbf{Fold3}
& \textbf{Fold4}
& \textbf{Fold5}
& \textbf{Average}
\\
\hline
    \multirow{4}{*}{\textbf{128}}
& \textbf{64}
& 83.82 ($\pm$3.22)
& 94.03 ($\pm$2.02)
& 94.58 ($\pm$1.9)
& 94.29 ($\pm$2.05)
& 90.84 ($\pm$2.58)
& 91.51 ($\pm$2.35)
\\
& \textbf{128}
& 83.98 ($\pm$3.17)
& 94.09 ($\pm$2.00)
& 94.82 ($\pm$1.86)
& 94.65 ($\pm$1.94)
& 90.89 ($\pm$2.45)
& 91.69 ($\pm$2.28)
\\
& \textbf{256}
& 84.74 ($\pm$3.13)
&94.60 ($\pm$1.92)
& 95.19 ($\pm$1.80)
& 95.06 ($\pm$1.86)
& 91.59 ($\pm$2.44)
& 92.24 ($\pm$2.23)
\\
        & \textbf{512}
& 85.27 ($\pm$3.12)
& 95.55 ($\pm$1.70)
& 95.81 ($\pm$1.65)
& 95.60 ($\pm$1.73)
& 92.16 ($\pm$2.32)
& 92.88 ($\pm$2.10)
\\
\hline
\hline
\end{tabular}}
}
\end{center}
\end{table}
\begin{table}[t!]
 \begin{center}
 \caption{\footnotesize The number of learnable parameters for different number of electrodes and window sizes.}
 \label{params}
 \resizebox{\columnwidth}{!}{
{\begin{tabular}{  c c c c c }
\hline
\hline
\textbf{\# Channels}
& \textbf{Window size}
& \textbf{\# Parameters ($\MM$-V1)}
& \textbf{\# Parameters ($\MM$-V2)}
& \textbf{\# Parameters (3D CNN)}
\\
\hline
    \multirow{3}{*}{\textbf{32}}
& \textbf{64}
& 46,530
& -
& -
\\
& \textbf{128}
& 47,042
    & -
    & -
\\
& \textbf{256}
& 48,066
        & -
        & -
\\
\hline
\multirow{3}{*}{\textbf{64}}
& \textbf{64}
&62,914
    & -
    & 294,914
\\
& \textbf{128}
&63,426
    & -
    & 311,298
\\
& \textbf{256}
& 64,450
    & -
    & 319,490
\\
\hline
\multirow{4}{*}{\textbf{128}}
& \textbf{64}
& 95,682
    & 273,346
    & -
\\
& \textbf{128}
& 96,194
    & 274,370
    & -
\\
& \textbf{256}
& 97,218
    &276,418
    & -
\\
& \textbf{512}
& 99,266
& 280,514
& -
\\
\hline
\hline
\end{tabular}}
}
\end{center}
\end{table}
In this experiment, we employ $4$ different window sizes together with $3$ different combination of electrodes of the HD-sEMG dataset and report the achieved accuracy for each of the $5$ test folds and the overall averaged accuracy.
In the first model, referred to as the $\MM$-V1, the simplest and smallest $\MM$ model that gives acceptable results is chosen. The length of windowed signals, in this model, is set to $64$, $128$, $256$ and $512$ ($31.25$, $62.5$, $125$, $250$ ms respectively) with skip step of $32$ except for the window size of $512$ for which the skip step is set to $64$. To measure effects of increasing the number of channels on the performance of the proposed architecture, we consider three different settings using all, half, and $1/4$ of the $128$ electrodes. In the half mode, electrodes of multiple of $2$ and in the $1/4$ mode, electrodes of multiple of $4$ were chosen. As stated previously, the number of horizontal electrode channels in the $\MM$'s input is $4$, $8$, and $16$ while the number of vertical channels is $8$. Regarding the hyperparameters of the model, the model's (embedding) dimension is $64$, and the patch size is set to ($8,4$), ($8,8$), and ($8,16$) for $32$, $64$, and $128$ number of channels, respectively.
The $\MM$-V1 model contains only $1$ transformer layer and $8$ heads. The MLP block's hidden size is set to $64$, the same as its input size. The $\MM$-V1 model is trained with $20$ epochs and batch size of $128$ for each subject independently. The optimization method used is Adam with $\beta1 = 0.9$ and $\beta2 = 0.999$ parameters, learning rate of $0.0001$ and weight decay of $0.001$. Learning rate annealing is deployed after the first $10$ epochs for faster convergence. The cross-entropy loss function is considered as the objective function. Table~\ref{model1} represents the acquired accuracy and standard deviation (STD) for each individual window size and number of channels. It is worth noting that the $512$ window size is only tested with the whole electrode channels of the dataset to indicate the potential best performance of the network.

A second variant of the $\MM$ model, referred to as $\MM$-V2, is also tested where the model's dimension and the number of hidden layers in the MLP layer are twice those of $\MM$-V1. We apply the $\MM$-V2 model on the data samples derived from the whole $128$ electrodes to compare it with the last $4$ rows of Table~\ref{model1}. The results are shown in Table~\ref{model2}. Table~\ref{params} illustrates the number of learnable parameters for each window size and number of channels in both models.

\begin{figure}[!t]
\centering
\includegraphics[scale=0.50]{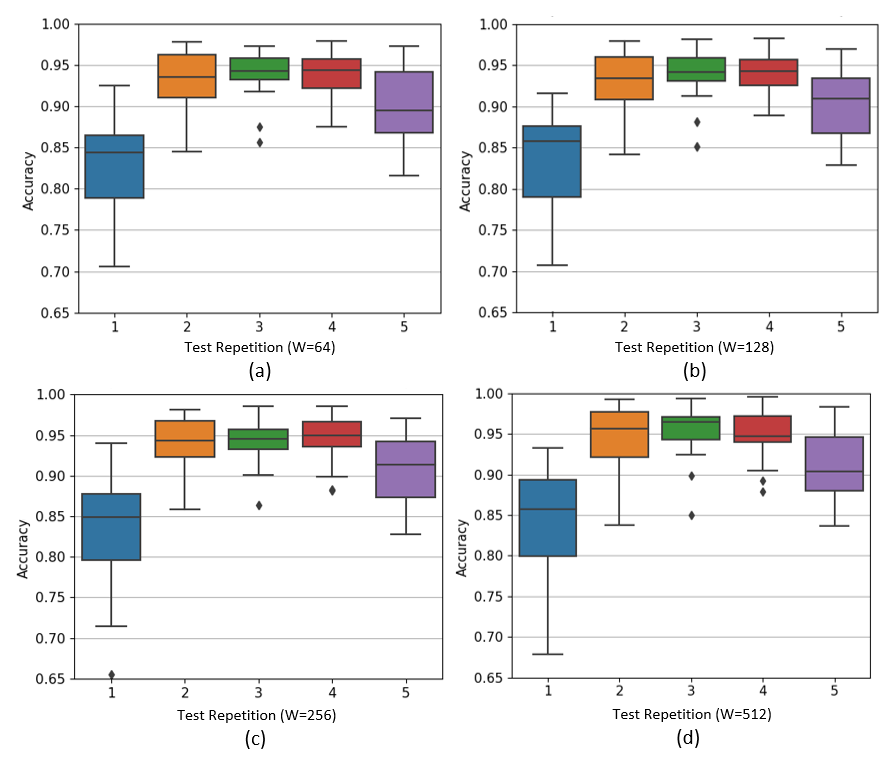}
\caption{\footnotesize Comparison of the accuracy $\MM$-V1 obtains for each fold and window sizes of (a) $W=64$ (b) $W=128$ (c) $W=256$ and (d) $W=512$. The number of utilized electrode channels in these plots is $128$.}
\label{repet}
\end{figure}

Fig.~\ref{repet} demonstrates the box plots for the accuracy of $\MM$-V1 obtained for each individual fold and different window sizes from $W=64$ to $W=512$ (Fig.~\ref{repet}(a-d)). The box plots are drawn based on the interquartile range (IQR) of accuracy for $19$ subjects when all the $128$ electrodes are included in the experiment. The black horizontal line represents the median accuracy for each fold.

\begin{figure}[t!]
\centering
\includegraphics[scale=0.45]{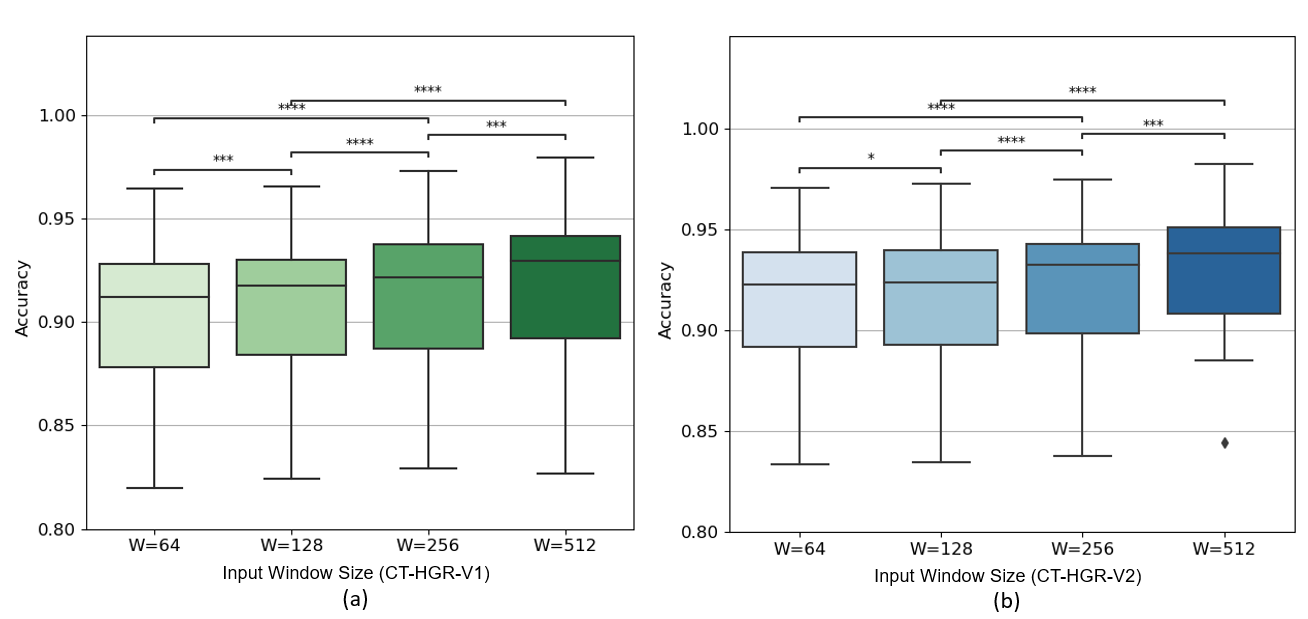}
\caption{\footnotesize Box plots and IQR for different window sizes ($W=64$,  $W=128$,  $W=256$, and  $W=512$) and $128$ number of channels for $\MM$- (a) V1 and (b) V2 }
\label{wilcoxon}
\end{figure}

In Fig.~\ref{wilcoxon}, the Wilcoxon signed rank test is applied for $\MM$-V1 and $\MM$-V2 separately when the number of channels is fixed at $128$. The box plots show the IQR for each window size that decreases minimally from $\MM$-V1 to $\MM$-V2. The Wilcoxon test's $p$-value annotations in Fig.~\ref{wilcoxon} are as follows:

\begin{itemize}
\item ns: $5.00e-02 < p <= 1.00e+00$
\item *: $1.00e-02 < p <= 5.00e-02$
\item **: $1.00e-03 < p <= 1.00e-02$
\item ***: $1.00e-04 < p <= 1.00e-03$
\item ****: $p <= 1.00e-04$
\end{itemize}
Although the average accuracy does not change significantly, the STD in $\MM$-V2 with $W=512$ declines significantly compared to $\MM$-V1.

\begin{figure}[t!]
\centering
\includegraphics[width=0.8\linewidth]{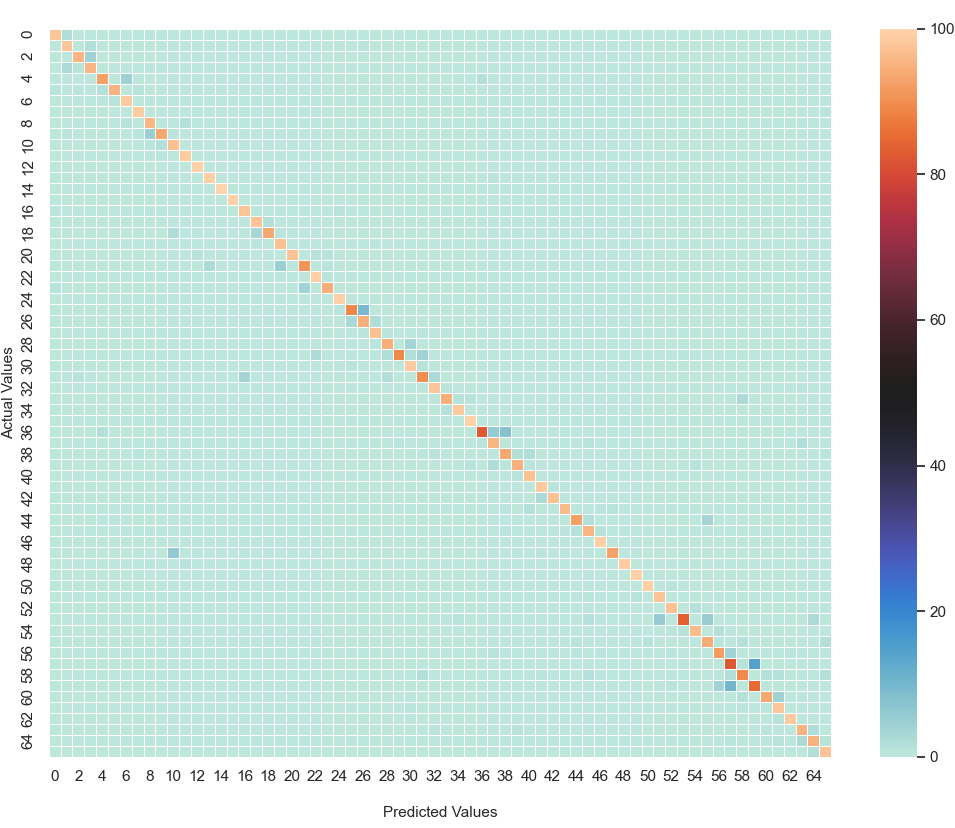}
\caption{\footnotesize Average confusion matrix of Model $\MM$-V1 with $W=512$ and $128$ number of electrodes over $19$ subjects. }
\label{conf}
\end{figure}
The gestures in the HD-sEMG dataset are ordered according to their DoF and similarity in performance. The simple 1 DoF gestures are labeled from $1$ to $16$, 2 DoF gestures are from $17$ to $57$ and the most complex ones are from $58$ to $66$. To be more specific, the confusion matrices for Model $\MM$-V1 with $W=512$ and $128$ number of channels are obtained for all the subjects. The matrices are summed and normalized row-wise. The final confusion matrix is shown in Fig.~\ref{conf}. The diagonal values show the average accuracy acquired for each hand gesture among $19$ subjects. The average accuracy for most of the gestures is above $94$\%. The density of the non-zero elements in Fig.~\ref{conf} is utmost near the diagonal, which implies that the possibility of the network making mistakes in gesture classification is higher in gestures that have the same DoF and are performed similarly.

\vspace{2cm}
\subsection{Comparisons with a Conventional ML and a 3D Convolutional Model}\label{sub:compare}

\begin{figure*}[t!]
\centering
\includegraphics[scale=0.40]{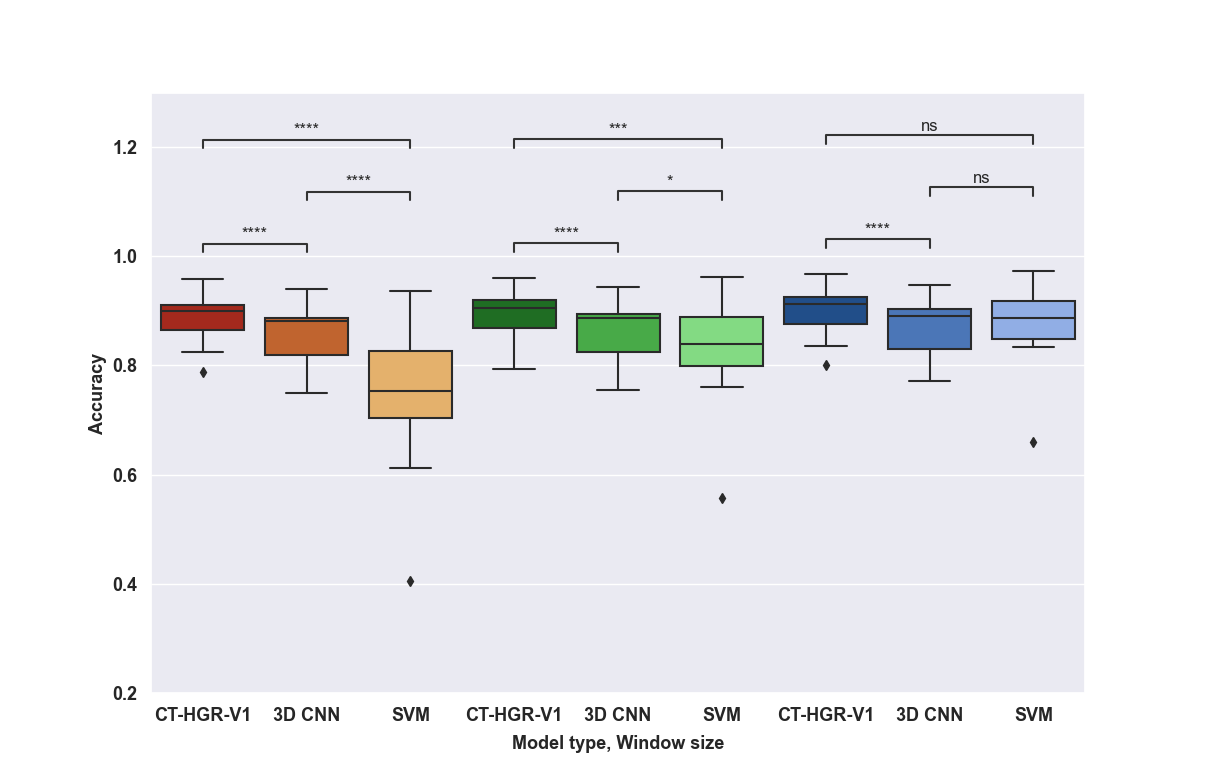}
\caption{\footnotesize Box plots and IQR of $\MM$-V1, 3D CNN and SVM for different window sizes ($W=64$  $W=128$ and $W=256$) and $64$ number of channels.}
\label{comp}
\end{figure*}

\begin{table}[t!]
 \begin{center}
\caption{\footnotesize Comparison of classification accuracy and STD of each fold and their average for different window sizes and $64$ electrode channels using an SVM model. The accuracy and STD for each fold is averaged over $19$ subjects.}
 \label{svm}
 \resizebox{\columnwidth}{!}{
 {\begin{tabular}{  c  c  c c c c c c}
\hline
\hline
\textbf{\# Channels}
& \textbf{Window size}
& \textbf{Fold1(\%)}
& \textbf{Fold2}
& \textbf{Fold3}
& \textbf{Fold4}
& \textbf{Fold5}
& \textbf{Average}
\\
\hline
    \multirow{3}{*}{\textbf{64}}
& \textbf{64}
& 63.62 ($\pm$11.58)
& 77.62 ($\pm$11.28)
& 78.99 ($\pm$11.45)
& 78.12 ($\pm$11.07)
& 73.42 ($\pm$11.10)
& 74.35 ($\pm$11.30)
\\
& \textbf{128}
& 72.84 ($\pm$10.47)
& 86.49 ($\pm$8.32)
& 87.32 ($\pm$8.11)
& 87.48 ($\pm$8.12)
& 82.35 ($\pm$8.66)
& 83.30 ($\pm$8.74)
\\
& \textbf{256}
& 77.74 ($\pm$9.34)
& 90.95 ($\pm$6.18)
& 91.56 ($\pm$6.06)
& 91.01 ($\pm$5.84)
& 87.41 ($\pm$6.51)
& 87.73 ($\pm$6.79)
\\
\hline
\hline
\end{tabular}}
}
\end{center}
\end{table}

\begin{table}[t!]
 \begin{center}
\caption{\footnotesize Comparison of classification accuracy and STD of each fold and their average for different window sizes and $64$ electrode channels using a 3D CNN model. The accuracy and STD for each fold is averaged over $19$ subjects.}
 \label{cnn}
 \resizebox{\columnwidth}{!}{
 {\begin{tabular}{  c  c  c c c c c c}
\hline
\hline
\textbf{\# Channels}
& \textbf{Window size}
& \textbf{Fold1(\%)}
& \textbf{Fold2}
& \textbf{Fold3}
& \textbf{Fold4}
& \textbf{Fold5}
& \textbf{Average}
\\
\hline
    \multirow{3}{*}{\textbf{64}}
& \textbf{64}
& 75.90 ($\pm$3.85)
& 89.19 ($\pm$2.65)
& 90.36 ($\pm$2.56)
& 89.81 ($\pm$2.63)
& 85.51 ($\pm$3.06)
& 86.15 ($\pm$2.95)
\\
& \textbf{128}
& 76.30 ($\pm$3.73)
& 90.13 ($\pm$2.56)
& 90.84 ($\pm$2.42)
& 90.06 ($\pm$2.57)
& 86.07 ($\pm$2.98)
& 86.68 ($\pm$2.85)
\\
& \textbf{256}
& 77.63 ($\pm$3.62)
& 90.56 ($\pm$2.49)
& 90.95 ($\pm$2.45)
& 90.95 ($\pm$2.45)
& 87.14 ($\pm$2.83)
& 87.45 ($\pm$2.77)
\\
\hline
\hline
\end{tabular}}
}
\end{center}
\end{table}

In the first part of this sub-section, we provide comparison results with a traditional ML algorithm developed based on Support Vector Machines (SVM), which is commonly~\cite{svm1,svm2,svm3} used for hand gesture recognition tasks. Following the works of References~\cite{svm1,svm2,svm3}, we extract the four most popular feature sets that are used for ML classifiers from our dataset. Extracted features are as follows: Root Mean Square (RMS), Zero Crossings (ZC), Slope Sign Change (SSC), and Wavelength (WL), which are separately extracted for each electrode channels. In Table~\ref{svm}, the obtained results for the SVM model in which the number of channels in the dataset is set to $64$ are presented.

In the second part, we implement a 3D CNN model that is originally utilized for video-based hand gesture recognition tasks~\cite{cnn3d-cvpr} and is found effective by authors in~\cite{cnn3d} to be applied on HD-sEMG datasets as they resemble video data in having one dimension in time and two dimensions in space. Therefore, in spite of a typical 2D CNN model, a 3D CNN architecture is able to extract both the temporal and spatial features in HD-sEMG datasets. The 3D signals of shape $W\times N_{ch}\times N_{cv}$ go through the 3D CNN architecture that has two consecutive 3D CNN layers with $16$ and $32$ respective filters of size ($5,3,3$), each followed by a GELU activation function, a dropout and a max pooling layer. Then, two fully connected (FC) layers of size $256$ and $128$ are deployed before the output layers which consists of an MLP head similar to the one used in our $\MM$ models followed by a $softmax$ function for classification. The other hyperparameters of the network are set similar to those of the $\MM$ model. The stride values in both 3D CNN layers are $1$. Table~\ref{cnn} shows the acquired results for the 3D CNN model in which the number of channels in the dataset is set to $64$.

Fig.~\ref{comp} shows the box plots and the results of Wilcoxon signed rank statistical test that is conducted for comparing $\MM$-V1 and SVM model's accuracy on $19$ subjects. In this experiment, only the models with the same window size are compared to assess the discrepancy between two different models with the same input data.

\subsection{Performance Evaluation based on Shuffled Data}\label{sub:shuffle}
\begin{table}[t!]
 \begin{center}
 \caption{\footnotesize Accuracy and STD for the shuffled dataset of all the $5$ repetitions and different window sizes ($\MM$-V1).}
 \label{shuffle}
    {\begin{tabular}{  c  c  c }
\hline
\hline
\textbf{\# Channels}
& \textbf{Window size}
& \textbf{\# Avg accuracy(\%)}
\\
\hline
    \multirow{3}{*}{\textbf{64}}
& \textbf{64}
& 98.05 ($\pm$1.19)
\\
& \textbf{128}
& 98.43 ($\pm$1.05)
\\
& \textbf{256}
& 98.79 ($\pm$0.96)
\\
\hline
\hline
\end{tabular}}
    \end{center}
\end{table}
In the previous sub-sections, a $5$-fold cross-validation was applied on the HD-sEMG dataset in which the test set (repetition) is entirely unseen and is not included in the train set (repetitions). However, another approach followed in the literature~\cite{ml2022,split} to split the two sets is to shuffle the whole dataset with $5$ repetitions and assign $20\%$ for the test set and $80\%$ for the train set. In this scenario, the test set is unseen but data samples from the same repetitions may exist in both sets making the model more familiar with the test samples potentially achieving higher accuracy. The obtained average accuracy over $19$ participants using $64$, $128$, $256$ window sizes using the hyperparameters of $\MM$-V1 are summarized in Table~\ref{shuffle}.

\subsection{Instantaneous Performance Evaluation}\label{sub:instant}
\begin{table}[t!]
\begin{center}
\caption{Accuracy and STD of each fold and their average for instantaneous training.}
\label{instant}
\resizebox{\columnwidth}{!}{
\begin{tabular}{c c c c c c c c}
\hline
\hline
\textbf{\# Channels}
& \textbf{Window size}
& \textbf{Fold1(\%)}
& \textbf{Fold2}
& \textbf{Fold3}
& \textbf{Fold4}
& \textbf{Fold5}
& \textbf{Average}\\
\hline
\textbf{64}
& \textbf{1}
& 80.02 ($\pm$3.45)
& 92.33 ($\pm$2.27)
& 92.47 ($\pm$2.26)
& 92.16 ($\pm$2.31)
& 88.69 ($\pm$2.74)
& 89.13 ($\pm$2.61) \\
\hline
\end{tabular}
}
\end{center}
\end{table}
In this sub-section, our objective is to assess the functionality of the proposed framework on instant HD-sEMG data points. In other words, we consider window size of only $1$ sample as the input to our model, which requires no patching. We set the number of electrodes to $64$. The hyperparameters used in this experiment are the same as those used for $\MM$-V1. The accuracy results are presented in Table~\ref{instant}.

\subsection {Evaluation of a Hybrid Model based on Raw HD-sEMG and Extracted MUAPs}\label{sub:hybrid}
\begin{figure}[t!]
\centering
\includegraphics[scale=0.5]{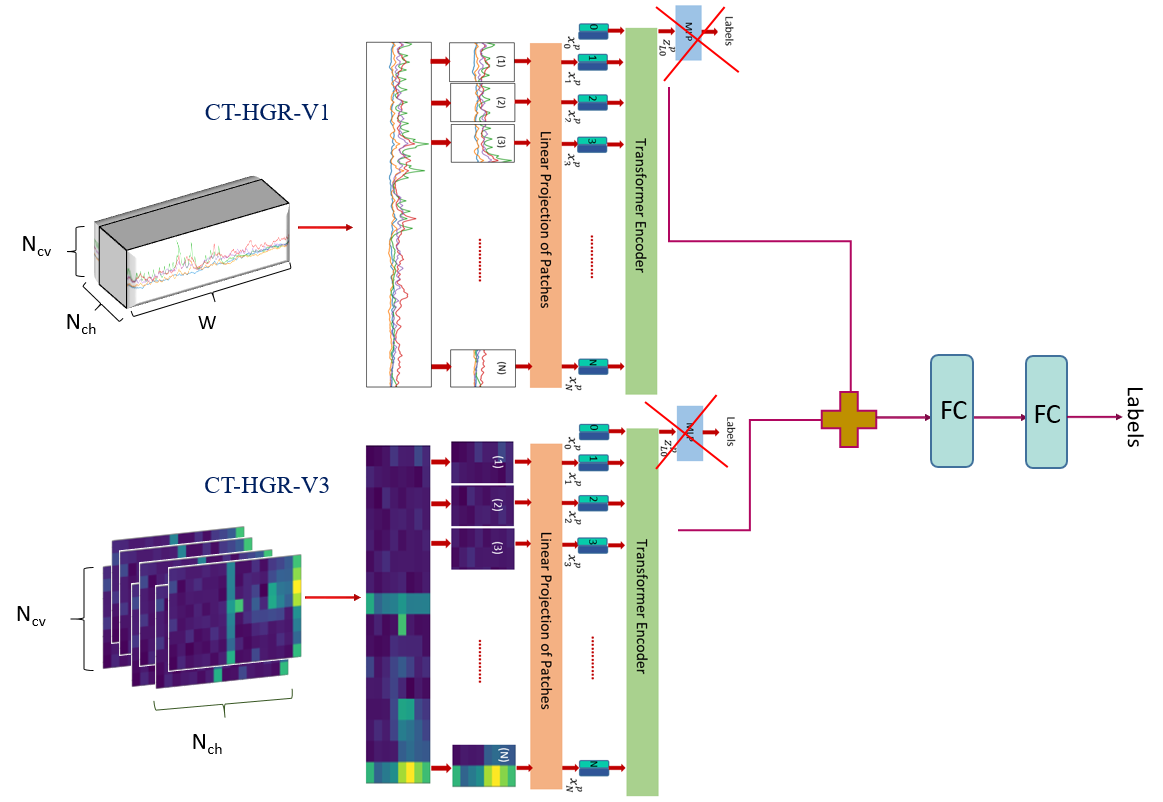}
\caption{\footnotesize The fused $\MM$ framework. In the first stage, the ViT-based models in  the Macro and Micro paths are trained based on 3D, HD-sEMG and 2D, p-to-p MUAP images, respectively. In the second stage, the Micro and Micro weights are frozen. The final Micro and Macro class tokens are concatenated and converted to a $1,024$-dimensional feature vector, which is fed to a series of FC layers for gesture classification. }
\label{hybrid}
\end{figure}
In this sub-section, we present the results of fusing $\MM$-V1 with a third variant of $\MM$ called $\MM$-V3 that works based on the extracted MUAPs from raw HD-sEMG signals. Fig.~\ref{hybrid} illustrates the overall hybrid architecture of the fused model. More specifically,  $\MM$-V3 uses HD-sEMG decomposition to extract microscopic neural drive information. sEMG decomposition refers to a set of Blind Source Separation (BSS)
methods that extract discharge timings of motor neuron action potentials from raw HD-sEMG data. Single motor neuron action potentials are summed to form MUAPs that convert neural drive information to hand movements.
Motor unit discharge timings, also known as Motor Unit Spike Trains (MUSTs), represent sparse estimations of the MU activation times with the same sampling frequency and time interval as the raw HD-sEMG signals.
HD-sEMG signals can be modelled as a spatio-temporal convolution of MUSTs, which provide an exact physiological description of how each hand movement is encoded at neurospinal level~.
Thus, MUSTs are of trustworthy and discernible information on the generation details of different hand gestures, as such is adopted in $\MM$-V3 for hand gesture recognition.  For extracting MUSTs, among the existing BSS approaches~\cite{bss} suggested for HD-sEMG decomposition, gradient Convolution Kernel Compensation (gCKC)~\cite{holobara,holobarb} and fast Independent Component Analysis (fastICA)~\cite{fastica} are of great prominence and frequently used in the literature. To achieve better accuracy, the utilized BSS algorithm~\cite{bss} is a combination of gCKC~\cite{holobara,holobarb} and fastICA~\cite{fastica} algorithms.  In this method, the number of extracted sources is dependent on the number of iterations in which a new MU is found and also the silhouette threshold that admits sources with high quality. Finally, a fused version is also designed to simultaneously extract a set of temporal and spatial features through its two independent ViT-based parallel paths. The \textit{Macro Path} is the $\MM$-V1, while the \textit{Micro path} is $\MM$-V3  fed with the peak-to-peak values of the extracted MUAPs of each source. A fusion path, structured in series to the parallel ones and consisting of FC layers, then combines extracted temporal and spatial features for final classification.

In our experiment, the number of iterations is set to $7$ and the silhouette measure is set to $0.92$, so a maximum of $7$ sources are estimated for each windowed signal. Considering multi-channel sEMG signals as a convolutive mixture of a set of impulse functions known as the MUSTs of each MU~\cite{bss}, MUSTs are estimated for each electrode channel separately. Therefore, each windowed signal of shape $W\times N_{ch}\times N_{cv}$ is of maximum $7$ MUs that retain various activation levels for each electrode channel. As stated in~\cite{zhao_muap,dario_muap}, the activation level/area of MUs in limb muscles is highly variable across different hand gestures. Accordingly, if the peak-to-peak values of MUAPs for each MU and all the channels is calculated, a set of 2D images are acquired which have a predictable pattern among different hand gestures. These 2D images are considered as new input data to the $\MM$-V3. After training $\MM$-V1 and V3 individually, the models' weights are kept constant, the final class tokens of each model are joined together and fed to an FC layer for final classification. In such a way, the hybrid model decides based on preprocessed HD-sEMG signals as well as peak-to-peak images of MUAPs obtained for each MU independently. The $\MM$-V3's hyperparameters are set as follows: For both $\MM$-V1 and V3, HD-sEMG data is divided into windows of shape ($512$,$8$,$16$) with skip step of $256$. So, the image size and the number of input channels for 2D images are set to ($8\times16$), $1$ respectively. For each peak-to-peak image, we considered $2$ patches by setting patch size as ($8\times8$). The model's embedding dimension ($\d$) and number of heads is the same as the two previous models. The optimization algorithm is Adam with learning rate of $0.0003$ and weight decay of $0.001$. Each batch has $64$ data samples and the model is trained through $50$ epochs. Table~\ref{comp2} compares accuracy and STD for $\MM$-V1, $\MM$-V3 and their fused model for each fold.

\begin{table*}[t]
 \begin{center}
 \caption{\footnotesize Comparison of classification accuracy and STD for each fold and their average for each of the $3$ models. The accuracy and STD for each fold is averaged over $19$ participants.}
 \label{comp2}
 {\begin{tabular}{c c c c c c c}
\hline

 \textbf{Model Name}
& \textbf{Fold1(\%)}
& \textbf{Fold2}
& \textbf{Fold3}
& \textbf{Fold4}
& \textbf{Fold5}
& \textbf{Average}
\\
\hline

{$\MM$-V1}
& 79.92 ($\pm$3.39)
& 91.43 ($\pm$2.48)
& 93.84 ($\pm$2.05)
& 92.57 ($\pm$2.28)
& 88.96 ($\pm$2.83)
& 89.34 ($\pm$2.61)
\\
{$\MM$-V3}
& 81.53 ($\pm$3.45)
& 88.03 ($\pm$2.66)
& 89.63 ($\pm$2.39)
& 89.11 ($\pm$4.02)
& 84.92 ($\pm$2.97)
& 86.64 ($\pm$3.10)
\\
 \textbf{Fused}
& \textbf{89.38 ($\pm$2.88)}
&\textbf{96.86 ($\pm$1.82)}
& \textbf{96.82 ($\pm$1.75)}
& \textbf{96.65 ($\pm$2.75)}
& \textbf{94.61 ($\pm$1.90)}
& \textbf{94.86 ($\pm$2.22)}
\\
\hline
\end{tabular}}
\end{center}
\end{table*}

\section{Discussion}\label{sec:discuss}
Based on the results shown in Table~\ref{model1} and Table~\ref{model2}, the accuracy for each fold and the average accuracy increases by increasing both the window size and the number of channels. Doubling the number of electrode channels from $32$ to $64$ results in $2-3$\%, and from $64$ to $128$ in $1-2$\% increase in all the reported accuracies. Generally speaking, increasing the window size feeds more data points to the model and helps the model to learn more effectively. Along the same line, choosing the same skip step (32 data points) for different window sizes expands the number of input samples to the model and leads to better generalization and higher accuracy. Generalization refers to the ability of the model to make correct predictions for previously unseen data samples. Feeding more data to the model results in more similarity between training and testing samples, hence forcing the network to generate more accurate predictions when new data is delivered. The small skip step (32) chosen here means that the predictions are made every $15.3$ms, causing a very small latency for real-time implementation of the proposed network in prosthetic devices. As it is evident from Table~\ref{model1}, starting from $86.23$\%, the average accuracy increases by $0.3-0.8$\% each time the window size is increased reaching $91.98$\% when the window size and the number of channels are at the maximum. Therefore, the number of utilized channels, in general, has a greater impact on the accuracy in comparison to the window size. Moreover, the smallest accuracy is for $Fold1$ while the highest is for $Fold3/Fold4$, which could be due to the fact that in the first repetition, the subject was not completely aware of the procedure and how to exactly perform the required gesture. Intuitively speaking, the subject was being trained to perform the requested task. We hypothesize that, in the \nth{3} and \nth{4} repetitions, the subject might have completely learned about the gesture and performed it more consistently, however, in the \nth{5} repetition, fatigue might be a factor resulting in lower performance and relatively large drop in the accuracy.
%
\begin{figure*}[t!]
\centering
\includegraphics[width=0.7\linewidth]{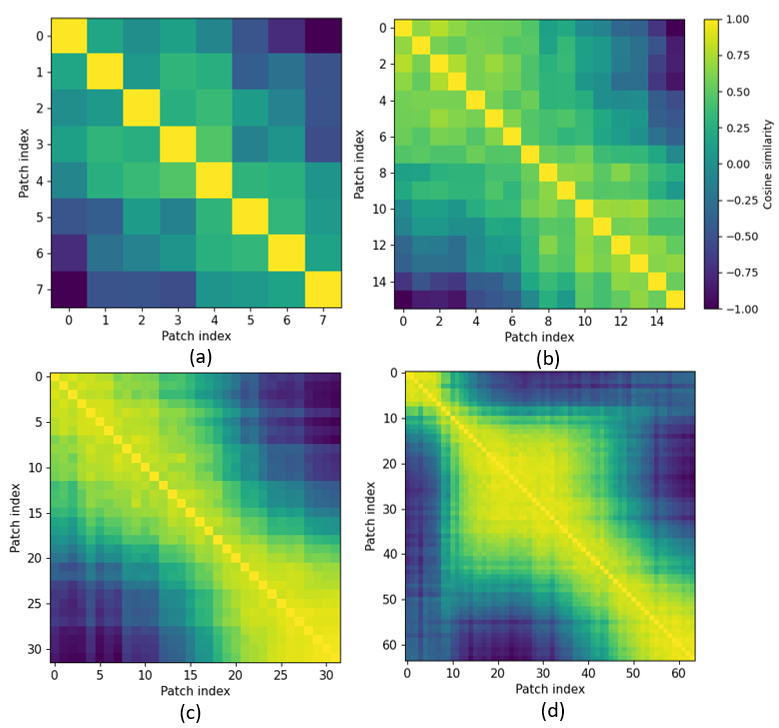}
\caption{\footnotesize Cosine similarities of repetition $3$, subject $20$ of $\MM$-V1 for (a) $W=64$ (b) $W=128$ (c) $W=256$ and (d) $W=512$}
\label{sim}
\end{figure*}
\begin{figure*}[t!]
\centering
\includegraphics[width=0.7\linewidth]{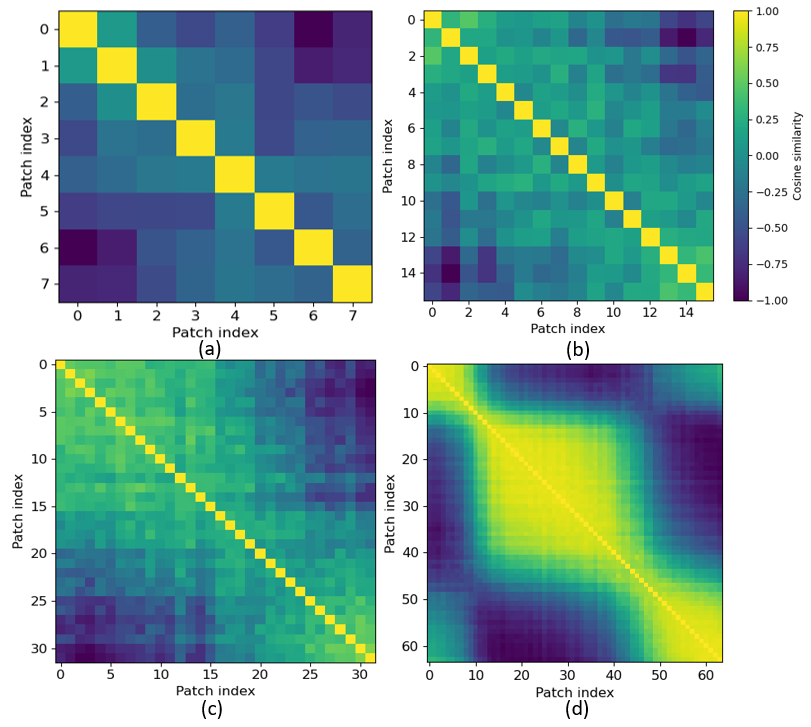}
\caption{\footnotesize Cosine similarities of repetition $3$, subject $20$ of $\MM$-V2 for (a) $W=64$ (b) $W=128$ (c) $W=256$ and (d) $W=512$}
\label{simi2}
\end{figure*}

As can be seen from Fig.~\ref{repet}, choosing the first repetition as the test set considerably differs from choosing the third or fourth repetition as the former yields much lower accuracy on average. STD for each fold and their average follows the same pattern as that of the accuracy, however, in an opposite direction, meaning that the best accuracy is usually associated with the least STD.

As can be seen in Table~\ref{model2}, Model $\MM$-V2 is generally a better model compared to its $\MM$-V1 variant as the accuracy for each fold and the overall average are higher. While the best improvement in accuracy occurs for $Fold1$ with $\approx$ $1.5$\% increase compared to $\MM$-V1, not much improvement (less than $1$\% in most cases) is observed  in the other folds and the final average. As indicated in Table~\ref{params}, $\MM$-V2's number of learnable parameters is roughly $3$ times the number of learnable parameters of $\MM$-V1, however, there is a marginal progress in its performance in comparison to the former model. This shows that the hyperparameters used in $\MM$-V1, producing no more than $100,000$ learnable parameters for the model, are sufficient for learning the $66$ hand movements with high accuracy and there is no need to use more complex models for hand gesture classification using the proposed $\MM$ framework on this specific HD-sEMG dataset. Clearly, deploying more complex models takes more memory and training time, which in turn reduces the overall efficiency of the model.
According to the box plots shown in Fig.~\ref{wilcoxon}, all the comparisons between different window sizes are statistically significant. For $\MM$-V2, we have $p \leq 0.001$ for the $W=64$ / $W=128$ and $W=256$ / $W=512$ pairs, which is less statistically significant than the other pairs with $p \leq 0.0001$. For $\MM$-V2, the results for the $W=64$ / $W=128$ pair are with $p \leq 0.05$ which is less statistically significant than that for the other pairs.

As mentioned previously, the positional embedding used in the $\MM$ framework is a $1$D trainable embedding vector that is added to each of the embedded patches. By increasing the window size in our experiments, the patch size remains constant and the number of patches increases. This causes the positional embedding, which is the principal factor in determination of the input samples' succession, to learn the positions more precisely. Fig.~\ref{sim} illustrates the cosine similarity matrices of the positional embedding in Model $\MM$-V1. Cosine similarities are sketched for different window sizes, $128$ electrode channels and the trained model on subject $20$ when repetition $3$ is considered as the test set. In this case, models with window sizes of $64$, $128$, $256$, and $512$ have ($8,16$) patch sizes. Therefore, each contain $8$, $16$, $32$ and $64$ patches in total. The $x$ and $y$ coordinates show the patch indices for each case and each row shows the similarities between each patch and the other patches. The diagonal values in each matrix are the largest values because their positional embedding vector is the same and its cosine is maximum. Similarity in the learned positional embedding vector of patches declines as the patches become farther. For $W=512$, the model learns the positions better and cosine similarities change more smoothly.
Fig.~\ref{simi2} demonstrates the cosine similarity matrices of the positional embedding in Model $\MM$-V2. Evidently, Model $\MM$-V2 has learned the position embeddings more effectively as there is less similarity between the distant patches for all the window sizes. The more the window size increases, the more the model discriminates between the distant patches and the more the adjacent patches are considered similar to each other. As illustrated in Fig.~\ref{sim} and Fig.~\ref{simi2}, for $W=512$, Model $\MM$-V2 behaves in a more orderly fashion than Model $\MM$-V1 and consequently, extracts the positional information better.

Regarding instantaneous training, authors in~\cite{butter} implemented a CNN to conduct instantaneous classification of $8$ gestures in the CapgMyo DB-a dataset. They applied various pre-processing and hyperparameter tuning steps and achieved a maximum accuracy of $88.1$\% when all the $128$ channels of the electrode grid were utilized. However, we achieved average accuracy of $89.13$\% with $64$ channels being involved in the study. Based on the results shown in Table~\ref{instant}, no significant discrepancy between the results for instantaneous training and larger window sizes is found. The results, in this case, are very similar to that of $\MM$-V1, when W=$128$ and number of channels is equal to $64$. This suggests that instantaneous training can sometimes work even better than training on very large window sizes with our proposed framework. More specifically, the model is able to achieve high accuracy in learning $66$ hand movements with a single-point input which can be considered as an important breakthrough in the field of hand gesture recognition. This proves that HD-sEMG datasets provide highly valuable information of the muscles' activity in each time point which are sufficient for the model to learn various hand gestures with no need for larger window sizes. Furthermore, training with single-point windows of data provides a great number of input samples to the $\MM$ which helps the model generalize better and avoid overfitting.
Based on the results shown in Table~\ref{shuffle}, the average accuracy and STD with shuffling is $\approx$ $9$\% higher and $\approx$ $1.4$\% lower than the results of the $5$-fold cross-validation, respectively. This, however, can cause major issues in practice when dealing with hand prosthetic devices since the test data is entirely unseen and the pre-trained model could not perform reliably while testing with new datasets. In other words, the results reported without shuffling should be used as the bases for practical utilization.

Based on the results shown in Table~\ref{svm} and Fig.~\ref{comp}, although increasing the window size leads to significant improvements in the average accuracy of the conventional SVM-based model, the achieved accuracy is still lower than the accuracy of the proposed $\MM$ architecture in all the cases. Furthermore, as indicated in Table~\ref{cnn} and Table~\ref{params}, our proposed $\MM$ framework surpasses the 3D CNN model by $\approx$ $3$\% average accuracy while employing less than $1/4$ of the learnable parameters used in the 3D CNN model. According to Table~\ref{cnn} and Fig.~\ref{comp}, the accuracy of both the deep networks ($\MM$-V1 and 3D CNN) increases by less than $1$\% with doubling the window size. As shown in Fig.~\ref{comp}, there is statistically significant difference among the three models with window size of $64$ ($p \leq 0.0001$), implying that the proposed $\MM$ gives its best performance at smaller window sizes. For $W=128$, the difference between $\MM$ and the other two models is still significant ($p \leq 0.0001$ and $p \leq 0.001$, respectively), but the 3D CNN and SVM do not differ considerably although the 3D CNN is of $\approx$ $3$\% more average accuracy than SVM. However, the proposed $\MM$ model seems to perform similarly to SVM when the window size is set to $256$ as the $p$-value in this case is $\geq 0.05$. In this case, there is still significant difference between $\MM$ and 3D CNN architectures with $p \leq 0.0001$.

According to Table~\ref{comp2} in which the studies are reported for the $250$ ms window size, $\MM$-V1's accuracy is higher than that of the $\MM$-V3 by $\approx$ $3-4$ \%, except $Fold1$ for which the peak-to-peak values of MUAPs provide more accurate information of the performed hand gesture than the HD-sEMG signals. However, a great improvement in average performance of the fused model in comparison to both stand-alone models is witnessed which is $8.22$ and $5.52$ \% increase compared to $\MM$-V1 and V3, respectively.

\section{Conclusion}\label{sec:conclu}
In this study, we proposed a ViT-based architecture, referred to as the $\MM$ framework, for hand gesture recognition from HD-sEMG signals. Efficacy of the proposed $\MM$ framework is validated through extensive set of experiments with various numbers of electrode channels and window sizes. Moreover, the proposed model is evaluated on instantaneous data samples of the input data, achieving, more or less, a similar accuracy to scenarios with larger window sizes. This provides the context for real-time learning from HD-sEMG signals. Although increasing the number of learnable parameters of the $\MM$ network leads to higher accuracy, the network works reasonably well on $66$ hand gestures with less than $65$k number of learnable parameters. This is exceptional as its conventional DL-based counterparts have, at times, millions of parameters. Besides, a hybrid model that is trained on raw HD-sEMG signals and their decomposed MUAPs is introduced, which substantially enhances the accuracy of the single $\MM$ model trained solely on raw HD-sEMG data. The sEMG decomposition method  utilized in this study, however,  is run offline which is considered as a drawback for active prosthetic systems that work in real time. As a direction for the future work, it is interesting to focus on real-time DL-based decomposition of sEMG signals rather than BSS with ICA or gCKC algorithms. In this case, the entire hybrid model can be used in an online fashion for incorporation in advanced HMI systems.
\bibliography{References}

\section*{Acknowledgments}
This Project was partially supported by Department of National Defence's Innovation for Defence Excellence \& Security (IDEaS), Canada, and Natural Sciences and Engineering Research Council (NSERC) of Canada through the NSERC Discovery Grant RGPIN 2019 06966.

\section*{Data Availability}
The utilized dataset is publicly available through the link provided in Reference~\cite{data}.

\section*{Author Contributions Statement}
M.M. and E.R. implemented the deep learning models and performed the evaluations; M.M and E.R. drafted the manuscript jointly with F.N. and A.M.;  F.A. and S.Y. contributed to the analysis and interpretation and edited the manuscript; F.N. and A.M. directed and supervised the study. All authors reviewed the manuscript.

\section*{Additional Information}
\textbf{Competing Interests}: Authors declare no competing interests.

\end{document}